\documentclass[%
 aip,
 amsmath,amssymb,
 reprint,%
reprint]{revtex4-1}

\usepackage{graphicx}
\usepackage{dcolumn}
\usepackage{bm}

\usepackage[utf8]{inputenc}
\usepackage[T1]{fontenc}
\usepackage{mathptmx}
\usepackage{etoolbox}
\usepackage{comment}
\usepackage{color}

\makeatletter
\def\@email#1#2{%
 \endgroup
 \patchcmd{\titleblock@produce}
  {\frontmatter@RRAPformat}
  {\frontmatter@RRAPformat{\produce@RRAP{*#1\href{mailto:#2}{#2}}}\frontmatter@RRAPformat}
  {}{}
}%
\makeatother
\begin{document}

\preprint{AIP/123-QED}

\title{High-Speed Tunable Generation of Random Number Distributions Using Actuated Perpendicular Magnetic Tunnel Junctions}
\author{Ahmed Sidi El Valli}
\affiliation{ 
Center for Quantum Phenomena, Department of Physics, New York University, New York, NY 10003 USA
}%

\author{Michael Tsao}%
\affiliation{ 
Center for Quantum Phenomena, Department of Physics, New York University, New York, NY 10003 USA
}%

\author{J. Darby Smith}
\affiliation{Neural Exploration and Research Laboratory, Sandia National Laboratories, Albuquerque,
New Mexico, United States. 
}%

\author{Shashank Misra}
\affiliation{Neural Exploration and Research Laboratory, Sandia National Laboratories, Albuquerque,
New Mexico, United States. 
}%

\author{Andrew D. Kent}
 \email{andy.kent@nyu.edu.}
\affiliation{ 
Center for Quantum Phenomena, Department of Physics, New York University, New York, NY 10003 USA
}%

\date{\today}%

\begin{abstract}
Perpendicular magnetic tunnel junctions (pMTJs) actuated by nanosecond pulses are emerging as promising devices for true random number generation (TRNG) due to their intrinsic stochastic behavior and high throughput. In this work, we study the tunability and quality of random-number distributions generated by pMTJs operating at a frequency of 104~MHz. First, changing the pulse amplitude is used to systematically vary the probability bias. The variance of the resulting bitstreams is shown to follow the expected binomial distribution. Second, the quality of uniform distributions of 8-bit random numbers generated with a probability bias of 0.5 is considered. A reduced chi-square analysis of this data shows that two XOR operations are necessary to achieve this distribution with p-values greater than 0.05. Finally, we show that there is a correlation between long-term probability bias variations and pMTJ resistance. These findings suggest that variations in the characteristics of the pMTJ underlie the observed variation of probability bias. Our results highlight the potential of stochastically actuated pMTJs for high-speed, tunable TRNG applications, showing the importance of the stability of pMTJs device characteristics in achieving reliable, long-term performance.
\end{abstract}

\maketitle



Today digital random numbers play a major role in advanced numerical computations, such as simulating stochastic processes  and artificial neural networks~\cite{kersevan2013monte,wang2017accelerating}. Most of these applications use pseudorandom numbers that are software-generated from a seed. However, a promising alternative is to use true random numbers harnessed from thermal fluctuations in nanoscale devices and converted to binary numbers. The latter is expected to offer higher-quality random numbers, simpler circuit architectures and better energy efficiency with increased throughput~\cite{misra2023probabilistic,borders2019integer}. 

One prominent candidate for random number generation is magnetic tunnel junction (MTJ) nanopillars, which consist of two ferromagnetic layers separated by a thin insulating barrier, patterned into a nanopillar structure. The relative orientation of the magnetizations of the two magnetic layers strongly influences the electric resistance of the junction~\cite{Kent2015,goldfarb2016introduction}. When the magnetizations are aligned parallel, the resistance is low; when they are antiparallel, the resistance is high. This dependence on magnetization alignment is a key feature that enables MTJs to convert thermally driven magnetic fluctuations into stochastic electrical signals.

The layer whose magnetization can be reoriented is called the free layer, while the other layer, with a fixed magnetization direction, serves as a reference. Random numbers can be generated by reducing the energy barrier to magnetization reversal of the free layer to a level where thermal noise drives its magnetization to randomly switch between parallel and antiparallel states~\cite{safranski2021demonstration,hayakawa2021nanosecond,soumah2024nanosecond}. This passive mechanism provides an energy-efficient approach for true random number generation (TRNG). However, the device's response is highly sensitive to applied fields and temperature variations~\cite{koh2024closed,rippard2011thermal}.

An alternative approach involves perpendicularly magnetized MTJs (pMTJs) with intermediate energy barriers, actuated by short electrical pulses that induce switching of the free layer's magnetization with a probability of approximately 50\%~\cite{rehm2023stochastic,rehm2024temperature,dubovskiy2024one}. These devices, termed stochastic magnetic-actuated random transducer perpendicularly magnetized MTJs (SMART-pMTJs), offer reduced sensitivity to temperature fluctuations~\cite{rehm2024temperature} and device-to-device variations~\cite{Morshed2023}, though at the cost of slightly higher energy consumption~\cite{rehm2023stochastic,shukla2023}.

In this article, we demonstrate the potential of SMART-pMTJs for high-speed and tunable TRNG applications. First, we investigate how varying the pulse amplitude enables control over the probability bias, confirming that the variance of the resulting bitstreams follows that expected of a Bernoulli process. Next, we evaluate the quality of uniform 8-bit random numbers generated with unbiased probability (probability bias of 0.5) using reduced chi-square tests. Finally, we identify a correlation between long-term variations in probability bias and resistance characteristics of pMTJ, suggesting that intrinsic variations in pMTJ properties drive these variations in probability bias. 

\begin{figure*}
    \centering
    \includegraphics[width=1.0\textwidth]{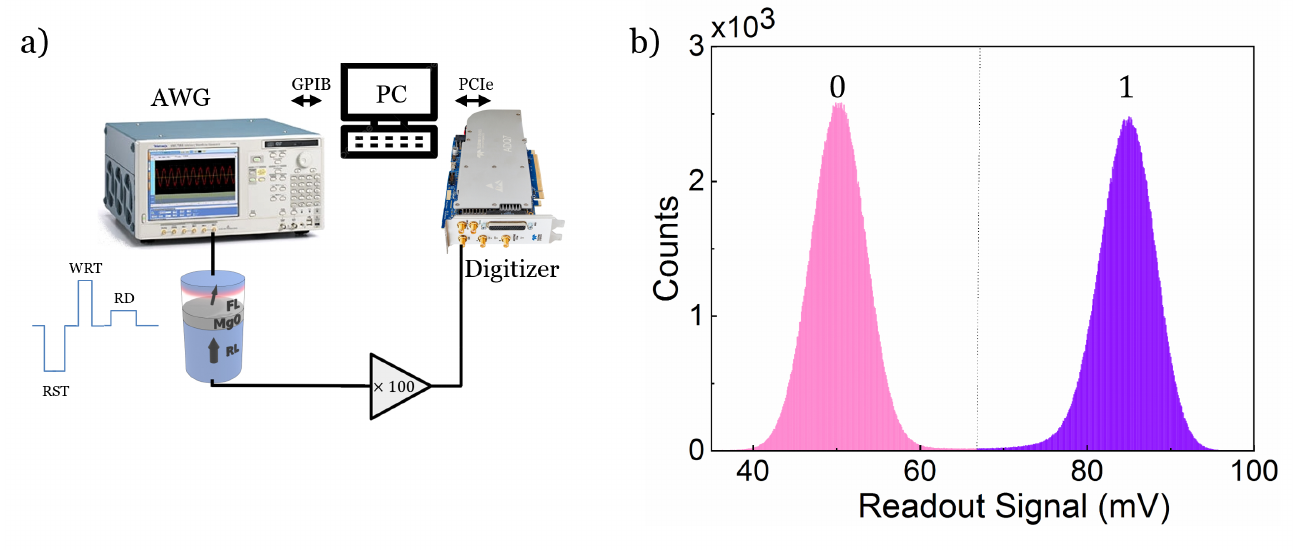}

    \caption{{\bf High-speed random number generation with SMART-pMTJ.} a) A schematic of the automated write/read experimental setup. An arbitrary waveform generator (10 GSPS Tektronix AWG7102) provides the voltage pulses that are applied to a pMTJ. A reset pulse (RST) sets the junction in the antiparallel state. This is followed by a stochastic write pulse (WRT) and then a readout pulse (RD). The full sequence is 9.6 ns in duration: 1ns RST + 2.6ns Idle + 1ns WRT + 0.4ns Idle + 4.6ns RD. The resulting output is amplified using a Tekronix low noise amplifier with a gain of 100 and digitized using a 14-bit Teledyne ADQ7DC digitizer. The signal is then processed within the Teledyne's internal FPGA. b) A histogram of the readout signal of $4.2\times 10^6$ random bits. The separation between the distributions of AP(0) and P(1) states is $\approx 6\sigma$, where $\sigma$ is the standard deviation of the distributions.}
    \label{Fig:Setup}
\end{figure*}

The experimental setup for generating high-speed true random bits using SMART-pMTJs is depicted in Fig.~\ref{Fig:Setup}(a). A sequence of three voltage pulses is applied to the pMTJ device using an arbitrary waveform generator (Tektronix AWG7102). The first pulse, labeled RST, initializes the device by resetting the free layer to a known state (the AP state in this case). This pulse has a sufficiently high amplitude to ensure deterministic switching. Next, a write pulse (WRT) with opposite polarity is applied to stochastically switch the free layer to the opposite state (the P state). The sequence concludes with a readout pulse (RD), which probes the state of the SMART-pMTJ after the WRT pulse. The RD pulse is low enough in amplitude to not alter the device state.

The timing of the sequence is as follows: the RST and WRT pulses are 1~ns in duration and separated by 2.6~ns, and the RD pulse, which lasts 4.6~ns, is applied 0.4~ns after the WRT pulse. The total duration of the sequence is thus 9.6~ns, enabling the generation of random bits at a rate of 104~Mb/s. The amplitudes of the RST and RD pulses are $0.92$ V and $0.2$ V, respectively. The experiments are performed on pMTJ nanopillars with a nominal diameter of 40~nm and an energy barrier of approximately $26kT$ for the AP$\rightarrow$P transition, where $k$ is the Boltzmann constant and $T$ is the device temperature of 300K~\citep{rehm2019sub}. The resistances of the AP and P states are $3.2\;\mathrm{k}\Omega$ and $2 \mathrm{k}\Omega$, respectively. Several devices were studied and typical results are reported here for one device.

To extract the random bits, the pMTJ output signal is first amplified using a Tektronix amplifier (gain: $100\times$). The amplified signal is then digitized by a high-performance 10GS/s digitizer (Teledyne ADQ7DC). The digitizer's internal FPGA processes the signal to isolate the readout pulse and identify the ``0'' (AP) or ``1'' (P) states of the SMART-pMTJ. Figure~\ref{Fig:Setup}(b) illustrates the histogram of readout signals obtained from approximately $10^6$ switching trials. The observed distributions for the AP state (low voltage) and P state (high voltage) are well-separated, with a voltage difference of approximately $6\sigma$, where $\sigma$ represents the standard deviation of the distributions.


\begin{figure*}
    \centering
    \includegraphics[width=1.0\textwidth]{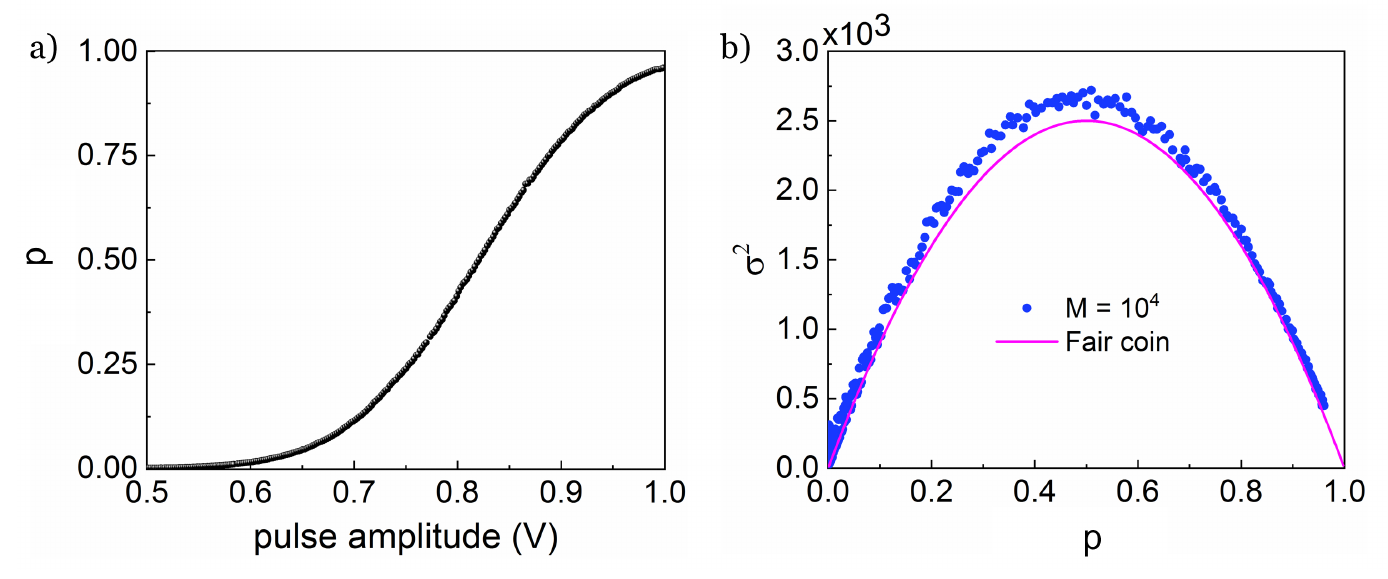}
    \caption{{\bf Switching probability analysis.} a) The measured AP$\rightarrow$P switching probability as a function of writing pulse amplitude. Each point corresponds to the average probability of 10 million independent trials. They were recorded at a speed of 104 Mb/s, and with a WRT duration of 1~ns. b) The variance of the switching probability is shown in (a) plotted in blue dots against the variance expected of a binomial distribution in pink. For every point, $10^4$ independent trials with $M=10^4$ switching attempts were considered.}
    \label{Fig:Sigmoid}
\end{figure*}
Figure~\ref{Fig:Sigmoid}(a) shows the switching probability for the AP$\rightarrow$P transition as a function of the amplitude of the stochastic write pulse. Each data point represents 10 million switching trials, acquired at a frequency of 104 Mb/s. The probability curve demonstrates precise tunability, with a vertical resolution of approximately $dp = 0.009$, determined by the 10-bit DAC resolution of the AWG ($\simeq 2\ \text{mV}$), and a maximum probability $p$ set by the highest pulse amplitude that can be applied while ensuring long-term endurance of the pMTJ.
 The observed sigmoidal shape aligns with the expected behavior in the macrospin ballistic regime, where spin-transfer torque dominates the switching dynamics~\cite{liu2014dynamics}, discussed and analyzed in Ref.~\cite{rehm2023stochastic}.

An essential criterion for evaluating tunable true random number sources is their fidelity to the binomial distribution~\cite{papoulis1990probability}. A binomial random variable is discrete and represents the number of successes (1's or P-state switching outcomes here), in $M$ trials. To assess this, we analyze the variance in the number of switching events, $N_s$, by grouping the 10 million trials for each pulse amplitude into $M = 10^4$ independent switching attempts. The variance is computed as $\sigma^2 = \langle N_s^2 \rangle - \langle N_s \rangle^2$, where the averages are taken over the independent switching trials. Figure~\ref{Fig:Sigmoid}(b) illustrates the variance as a function of the probability bias. The probabilities on the x-axis span the full range of the sigmoid curve shown in Fig.~\ref{Fig:Sigmoid}(a). The experimental results closely match the binomial variance formula, $\sigma^2 = M p (1 - p)$. Small deviations near $p = 0.5$ are attributed to the increased sensitivity to bias at the inflection point of the sigmoid curve.

\begin{figure*}[t]
    \centering
    \includegraphics[width=1.0\textwidth]{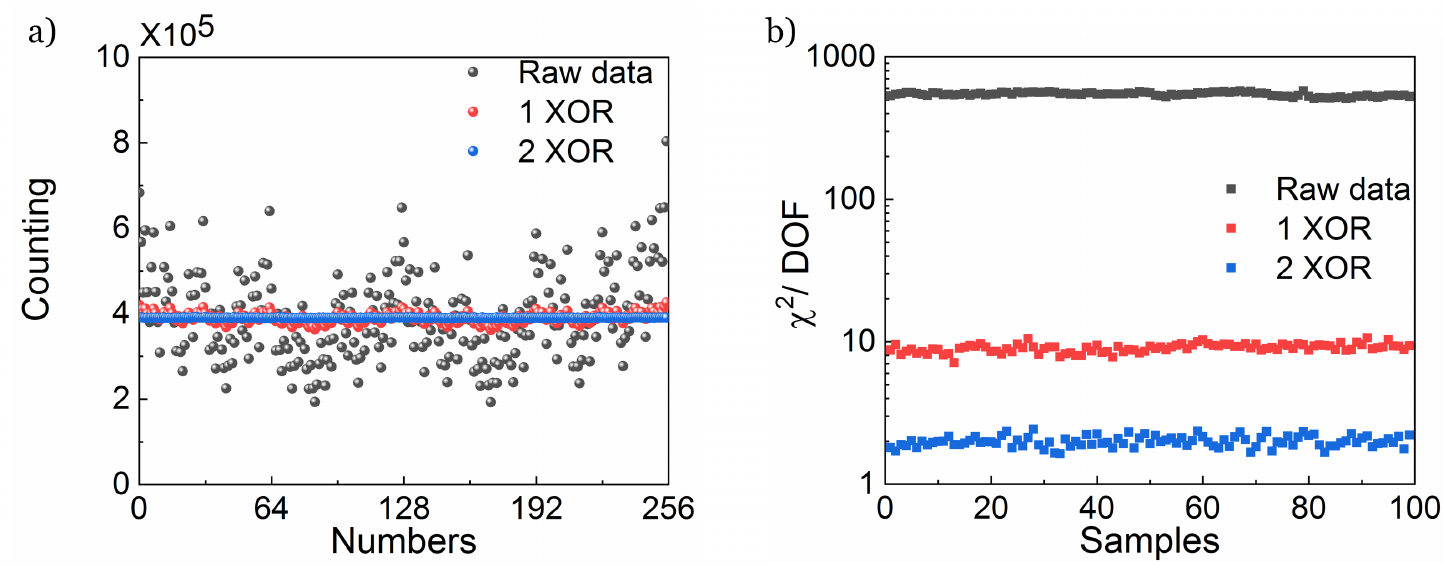}
    \caption{{\bf Statistical analysis of uniform sample.} a) Random bytes in integer format from the raw data (black), 1 XOR data (red) , and 2 XOR data (blue). b) The reduced chi-square test of uniform samples. Each sample consists of 1 million bytes (8 million bits). The sample numbers map their position in the recorded 100 million bytes. Raw SMART-pMTJ data is plotted in gray dots, XORed data is in red dots, and double XORed samples are in blue dots.}
    \label{Fig:Chisquare}
\end{figure*}

A fundamental operation in probabilistic computing applications is the ability to draw samples from various probability distributions. The suitability of different devices for such tasks can be assessed by evaluating the statistical quality of the samples generated by their bit streams. Here, we construct a nominally uniform distribution of 8-bit random numbers by sequentially placing eight consecutive readings of the SMART-pMTJ device, operating at a probability bias of $\approx 0.5$, in each position of an 8-bit string. The results are interpreted as a distribution of numbers ranging from 0 to 255. The resulting distribution is shown in Fig.~\ref{Fig:Chisquare}(a) as black points. It is known that exclusive OR (XOR) operations can be used to improve the quality of random bit streams~\cite{Matsui}. We XORed the data by splitting the full bit stream in half and XORing the streams bit by bit. This procedure was repeated to generate double-XORed (2 XOR) data. The XOR results are shown as the red (1 XOR) and blue (2 XOR) points in Fig.~\ref{Fig:Chisquare}(a). We note that the raw data exhibits noticeable deviations from a uniform distribution that are significantly reduced after the XORing operations.

To statistically analyze these results and the effect of XORing the data, we use the chi-square ($\chi^2$) test~\cite{papoulis1990probability}, which quantifies the deviation of the observed data from a uniform distribution. The reduced chi-square ($\bar{\chi}^2$) statistic is defined as:
\begin{equation}
\overline{\chi}^2 = \frac{1}{DOF} \sum_{i=1}^N \frac{(O_i - E_i)^2}{E_i},
\end{equation}
where $N = 256$ is the number of categories, $O_i$ is the observed count in the $i$-th category, $E_i$ is the expected count for a uniform distribution, and $DOF$ is the number of degrees of freedom, given by $DOF = N - 1$. A value of $\bar{\chi}^2 \approx 1$ indicates statistical consistency with the expected uniform distribution.

The results are shown in Fig.~\ref{Fig:Chisquare}(b) for a bitstream of $10^8$ bytes. The bitstream was divided into 100 samples of $10^6$ bytes each, and the chi-square test was performed on each sample individually. The raw data yields a $\bar{\chi}^2 \approx 560$, indicating a significant statistical deviation from the expected uniform distribution. After one XOR operation, $\bar{\chi}^2$ decreased to approximately 8, and following a second XOR operation, it further decreased to approximately 1.9. This demonstrates that two consecutive XOR operations are necessary to achieve statistical consistency with a uniform distribution. These conclusions were further verified using the p-value approach, with a passage threshold of 5\%, and the NIST statistical tests, which provide a comprehensive assessment of randomness~\cite{bassham2010statistical}, as summarized in Table~\ref{tab:table1}.

\begin{table}
\caption{\label{tab:table1} Evaluation of the random bits presented in Fig.~\ref{Fig:Chisquare}. The average p-values were computed using the full ${\chi}^2$ test (not normalized by the DOF). The NIST test results were graded based on the passage of the 16 available tests. For both p-values and NIST tests, 100 bins of $10^6$ random bits each were used.}
\begin{ruledtabular}
\begin{tabular}{ccd}
Samples & p-value & \text{NIST}\;\text{Tests}\\
\hline
Raw data & 0.0 & 0/16 \\
1 XORed data & 0.0 & 11/16\\
2 XORed data
  & 0.494 & 16/16 \\
\end{tabular} 
\end{ruledtabular}
\end{table}

\begin{figure}
    \centering
    \includegraphics[width=0.5\textwidth]{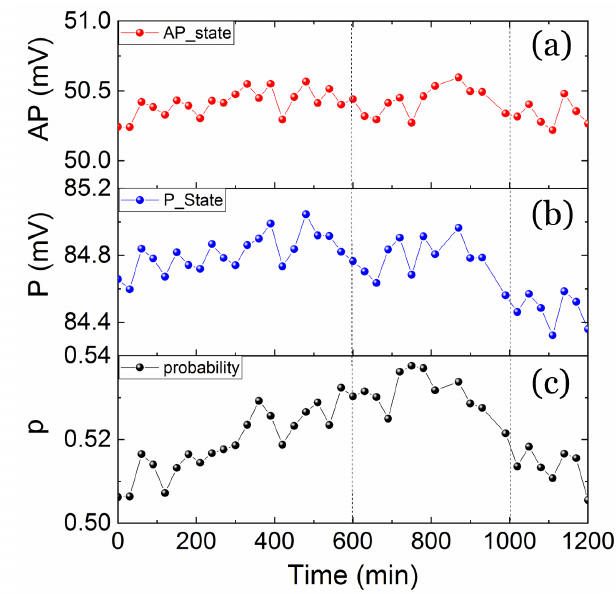}
    \caption{{\bf Spontaneous time measurements of the P/AP states, and the Probability.} a) The average AP-state voltages extracted from $10^8/32$ TRNG bit-stream. b) The average P-state voltages from the same bitstream in (a). c) The average probability of $10^8$ TRNGs recorded every 30 minutes for 1200 min. The dashed lines mark a measurement pause of 9 hours.}
    \label{Fig:deviation}
\end{figure}

Finally, to examine the possible origin of fluctuations in the raw data, we recorded the average probability bias of $10^8$ random bits every 30 minutes over a period of 20 hours. The results are shown in Fig.~\ref{Fig:deviation}. Dashed vertical lines indicate 9-hour pauses in the measurements, which were used to assess whether the characteristics of the device change over time when the device is not actuated. The relative continuity of the curves show that the changes in probability bias are correlated with the actuation events rather than the experiment time. The results in  Fig.~\ref{Fig:deviation}(c) indicate an initial increase in probability bias with a maximum deviation of approximately $3\%$, followed by a decrease toward $0.5$. Notably, the average parallel state (P state) resistance observed during the same measurements exhibits a similar temporal trend; see Fig.~\ref{Fig:deviation}(b). These observations suggest that fluctuations in the device's static resistance are correlated with variations in the probability bias, indicating that pMTJ characteristics evolve during electrical actuation.\\

These results highlight that improving the stability of the static resistance in pMTJs is key to enhancing the quality of the raw TRNGs. Further, as demonstrated in this study, incorporating just two XOR operations significantly enhances TRNG quality, making the bitstreams far less sensitive to variations in the device's probability bias.


In summary, we have demonstrated the potential of SMART-pMTJs for high-speed, tunable TRNG applications. Using nanosecond electrical pulses, we showed that the probability bias of these devices could be precisely controlled. In addition, we demonstrated that XOR operations are effective in improving the statistical quality of the generated random number distributions. Specifically, two consecutive XOR operations were required to achieve statistical consistency with a uniform distribution, as confirmed by chi-square testing, p-value analysis, and the NIST statistical randomness tests. 

Our investigation into the long-term stability of the SMART-pMTJ revealed a correlation between the device’s probability bias and its static resistance, suggesting that electrical actuation induces temporal variations in pMTJ characteristics. These findings underscore the importance of optimizing pMTJ stability to ensure consistent and reliable performance in TRNG applications. With their intrinsic stochastic behavior, high throughput, and tunability, SMART-pMTJs represent a promising technology for energy-efficient and scalable random number generation, with potential applications in cryptography, probabilistic computing, and large-scale Monte Carlo simulations. Further research into device-level optimization and integration with existing systems will enable broader adoption in next-generation computing architectures.
\begin{acknowledgments}
We acknowledge support from the DOE Office of Science (ASCR / BES) Microelectronics Co-Design project COINFLIPS and the Office of Naval Research (ONR) under Award No. N00014-23-1-2771. We thank Jonathan Z. Sun and Laura Rehm for helpful discussions of this work.

This paper describes objective technical results and analysis. Any subjective views or opinions that might be expressed in the paper do not necessarily represent the views of the U.S. Department of Energy or the United States Government. Sandia National Laboratories is a multimission laboratory managed and operated by National Technology \& Engineering Solutions of Sandia, LLC, a wholly owned subsidiary of Honeywell International Inc., for the U.S. Department of Energy’s National Nuclear Security Administration under contract DE-NA0003525.
\end{acknowledgments}

\section*{Data Availability Statement}
The data that supports the findings of this study are available from the corresponding authors upon reasonable request.
\nocite{*}
\bibliography{sMTJ.bib}

\end{document}